\documentclass[twocolumn,preprintnumbers,amsmath,amssymb,showpacs,superscriptaddress,footinbib,prl,aps,10pt]{revtex4-1}

\usepackage{graphicx}
\usepackage{bm}
\usepackage{dcolumn}
\usepackage{amssymb}
\usepackage{amsmath}
\usepackage{amsfonts}
\usepackage{epsfig}
\usepackage{graphicx}
\usepackage{stackrel}
\usepackage{paralist}
\usepackage[rgb]{xcolor}
\usepackage{subcaption}
\usepackage{todonotes}
\usepackage{pdfcomment}

\newcommand{\refEq}[1]{Eq.~(\ref{#1})}
\newcommand{\refFig}[1]{Fig.~\ref{#1}}
\newcommand{\refTab}[1]{Tab.~\ref{#1}}
\newcommand{\citeRef}[1]{Ref.~\onlinecite{#1}}

\begin{document}
\title{Coexisting edge states and gapless bulk in topological states of matter}
\author{Yuval Baum}
\affiliation{Department of Condensed Matter Physics, Weizmann Institute of Science, Rehovot 76100, Israel}
\author{Thore Posske}
\affiliation{Institut f{\"u}r Theoretische Physik und Astrophysik, Universit{\"a}t W{\"u}rzburg, 97074 W{\"u}rzburg, Germany}
\author{Ion Cosma Fulga}
\affiliation{Department of Condensed Matter Physics, Weizmann Institute of Science, Rehovot 76100, Israel}
\author{Bj{\"o}rn Trauzettel}
\affiliation{Institut f{\"u}r Theoretische Physik und Astrophysik, Universit{\"a}t W{\"u}rzburg, 97074 W{\"u}rzburg, Germany}
\author{Ady Stern}
\affiliation{Department of Condensed Matter Physics, Weizmann Institute of Science, Rehovot 76100, Israel}

\begin{abstract}
We consider two dimensional systems in which edge states coexist with a gapless bulk. Such systems may be constructed, for example, by coupling a gapped two dimensional state of matter that carries edge states to a gapless two dimensional system in which the spectrum is composed of a number of Dirac cones. We find that in the absence of disorder the edge states could be protected even when the two systems are coupled, due to momentum and energy conservation. We distinguish between weak and strong edge states by the level of their mixing with the bulk. In the presence of disorder, the edge states may be stabilized when the bulk is localized  or destabilized when the bulk is metallic. We analyze the conditions under which these two cases occur. Finally, we propose a concrete physical realization for one of our models on the basis of bilayer Hg(Cd)Te quantum wells.
\end{abstract}
\maketitle

\textit{Introduction ---}
The classification and realization of topological states of matter are among the main themes in modern condensed matter physics \cite{class,Kitaev,Zhang}. Of particular interest are topological insulators and topological superconductors, which have drawn a great deal of attention over the past few years \cite{HaldaneModel,mele,FKM,moore,BHZModel,Fukane,hassan,koning,Chen,Hsieh,Hsieh2}.
The bulk of these states is gapped, but the edge is commonly gapless. The gapless edge modes are protected from back-scattering and localization either by chirality or by symmetry. 

In this work, we present several examples of two dimensional models that simultaneously host one-dimensional gapless modes on the edge and are gapless in the two-dimensional bulk. In the absence of disorder, these models all share similar spectral and transport characteristics, showing distinct bulk and edge contributions that do not mix. In contrast, the effect of disorder unravels the difference between these models. In some of the models disorder stabilizes the edge modes and decouples them from the bulk, while in others it completely mixes the two. Similarly, the different nature of the different models is unraveled when we introduce weak perturbations that open energy gaps in the spectrum. 
We provide concrete Hamiltonians for the different models based on coupling a gapless phase to a topological phase, for example in a bilayer system, and analyze the spectral and transport properties of these models.
Beyond that, we present a concrete physical realization based on a bilayer Hg(Cd)Te quantum well. Note that the systems we consider share the lack of bulk energy gaps with Weyl semi-metals, but are distinguished from the latter by being two dimensional and by having no topological protection to the gapless nature of the bulk. 

The edge properties in gapped topological states of matter can be studied through the local density of states (LDOS). At energies smaller than the bulk energy gap, the LDOS is non-zero at the edge, and decreases exponentially as a function of the distance from the edge. Due to the absence of a bulk gap, this is not the case for the systems we consider, and we therefore have to employ different methods for studying the edge. For the clean case, we study a cylindrical geometry, in which the lattice momentum parallel to the edge, $k_{||}$, commutes with the Hamiltonian. We find two types of edge states, which we call strong and weak. Strong edge states carry a momentum $k_{||}$ and an energy $\epsilon(k_{||})$ for which there are no bulk states. Consequently, their wave functions are exponentially localized near the edge with the localization length being inversely proportional to the bulk gap at $k_{||}$. Weak edge states occur when for all values of $k_{||}$ and energies $\epsilon(k_{||})$ for which there are states at the edge, there are also states in the bulk. The edge states then hybridize with the bulk states, and their wave functions are not exponentially localized. We find that the ``strength'' of the edge mode depends on the orientation of the edge. For the models we consider, there are only a few orientations in which the edges are weak.  

The edge states are also reflected in transport. We distinguish between edge versus bulk transport by studying transport in devices of two terminals with both periodic and hard wall boundary conditions. This method is useful also in the presence of disorder, where states are not characterized by momentum.
Generally, in gapped phases with non-trivial topological index, edge state transport is robust as long as the relevant energy scales for transport are smaller than the bulk gap. Remarkably, for the systems we consider, the bulk gap vanishes, and yet the edge state transport may still be robust. In particular, we find that disorder may even stabilize the edge state transport.

The effect of disorder on the systems we consider may be inferred from the effect of a translational invariant perturbation that opens a gap in the bulk spectrum. Such a perturbation makes the system acquire a well-defined topological index. When the topological index is a Chern number whose value does not depend on the perturbation that opens the gap, the gapless phase is a transition between two insulating phases with the same Chern number. Then, in the presence of disorder, the bulk states may become localized and the edge states stabilize.
In contrast, when the gapless phase separates two topologically distinct insulating phases the phase diagram in the space of disorder and gap-opening perturbation must contain a critical line, where the bulk states remain delocalized and the edge states disappear. Away from the critical line, the system is a well-defined Chern insulator.
For cases where the topological index is not a Chern number, the localization properties of the bulk and edge states depend on the symmetries of the specific model.

We now introduce four models that share the same behavior in the absence of disorder, namely a coexistence of edge states and gapless bulk, but strongly differ away from that point. The models we consider are based on a designed coupling between a gapped two-dimensional topological phase $H_1$ and a gapless two-dimensional phase $H_2$. The simplest example would be a bilayer system in which the two layers are described by the Hamiltonians $H_1,H_2$ and are tunnel-coupled by $H_c$. The combined Hamiltonian can then be written as
\begin{equation} \label{BlockHamiltonian}
H=\begin{pmatrix}
H_1 & H_c \\
H_c^{\dagger} & H_2
\end{pmatrix}.
\end{equation}
Here, the topological phase $H_1$ is an insulator or a superconductor with a non-trivial topological index. For simplicity, we assume that the gap of $H_1$ is the largest energy scale. The Hamiltonian $H_2$ is gapless, for example, having a Dirac spectrum.
The coupling $H_c$ is chosen such that the full Hamiltonian remains gapless.
The different blocks should be combined such that the full Hamiltonian is irreducible, and hence it belongs to a symmetry class according to \citeRef{class}.
In general, the symmetry class of the full Hamiltonian is the minimal symmetry of $H_1$ and $H_2$, although, by fine tuning parameters, the resulting Hamiltonian may accidentally obey additional symmetries.

In the first three models $H_1$ describes a quantum Hall state with a non-zero Chern number, while $H_2$ describes three gapless phases that follow three different symmetries. As a consequence of the different symmetries, the effect of disorder on the three systems is markedly different. The fourth model belongs to a different topological class, but has the advantage of being experimentally accessible in a bilayer Hg(Cd)Te quantum well.

\textit{\label{QAH_4DCmodel}Model I --- A gapped Chern insulator coupled to a two-dimensional Dirac metal}. For the topological part, $H_1$, we take the Qi-Wu-Zhang Hamiltonian \cite{QAHE} of the quantum anomalous Hall effect,
\begin{align}\label{AQH}
H_1=&\epsilon(\textbf{k})-(t_0(\cos{k_x}+\cos{k_y})-\mu)\sigma_z \\ \nonumber
&+v_1(\sigma_x\sin{k_x}+\sigma_y\sin{k_y}),
\end{align}
where $\epsilon(\textbf{k})=t_1(1 -\cos{k_x}-\cos{k_y})\sigma_0$ is the kinetic energy and the $\sigma$'s are the Pauli matrices in spin space. Here and in the following, we set the lattice constant to $a=1$, as well as $t_0=1$, expressing all other Hamiltonian parameters relative to these scales. The model belongs to symmetry class A and has a non-zero Chern number for $0<\mu<2$.
For the gapless part, we use
$
H_2=v_2(\sigma_x\cos{k_x}+\sigma_z\sin{k_y}),
$
which contains four Dirac cones in its spectrum, at $(\pm\pi/2,0)$ and $(\pm\pi/2,\pi)$.
Notice that this model obeys effective time-reversal, particle-hole and chiral symmetries which all three square to unity: ${\cal T}=\sigma_x{\cal K}$, ${\cal P}=\sigma_z{\cal K}$ and ${\cal C}=\sigma_y$, respectively. Therefore, it belongs to class BDI, which is topologically trivial in two spatial dimensions.
Finally, we take $H_c=t(\sigma_0+\sigma_x)$ for the coupling Hamiltonian. In fact, for any $H_c$ with a zero determinant, the full Hamiltonian remains gapless.
Solving for its spectrum in a cylindrical geometry, we find weak edges when the boundary is along the $y$ direction or spans an angle of $\pm\arctan{(0.5)}$ with the $x$ axis. The strong edge states appear for all other boundary orientations.

\begin{figure}
\begin{center}
\centering
\includegraphics[width = \linewidth]{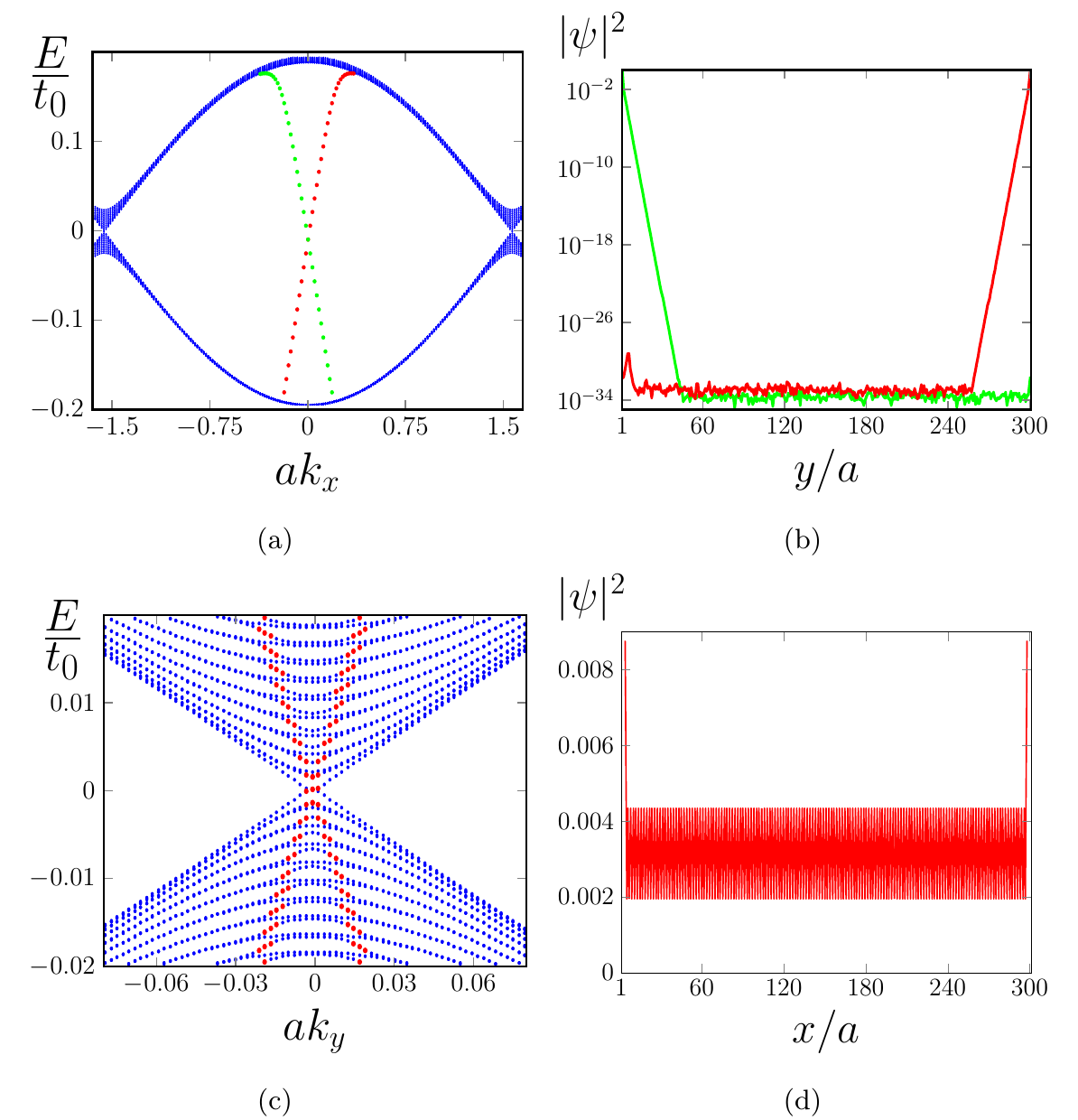}
\caption{(Color online) Band structure and edge state wave functions of \textit{model I} for different edge orientations. Edges along the y direction in (a) and (b): Well-localized states on the two edges (red, green) coexist with zero modes in the bulk (blue). Edges along the x direction in (c) and (d): Hybridized edge states (red) coexist with the bulk states.\label{fig:openYX}}
\end{center}
\end{figure}

The band structure of the system with open boundary conditions in the $y$ direction is shown in \refFig{fig:openYX}a. The blue points denote bulk states while the red/green points denote right/left propagating edge states, whose wave functions decay exponentially into the bulk (\refFig{fig:openYX}b). The zero modes in the bulk coexist with well localized chiral edge states. In contrast, the spectrum and typical edge states of a system with open boundary conditions in the $x$ direction are shown in \refFig{fig:openYX}c, d. Here, the red points denote edge states that hybridize with the bulk. We find numerically that the local density of states near the edge is larger than in the bulk (not shown), but the latter does not decay to zero at large distances from the edge.

Adding a mass term $m\sigma_y$ to $H_2$ opens a bulk gap. This term breaks both the $\cal P$ and $\cal C$ symmetries of $H_2$, but leaves $\cal T$ intact. Hence, the gapped version of $H_2$ belongs to class AI, which is also topologically trivial in two spatial dimensions \cite{class}. The full Hamiltonian, for $v_1\neq v_2$, is then a class A Chern insulator with a non-zero Chern number that is independent of the sign of $m$, so it belongs to the first class of models mentioned in the introduction.
Therefore, the system has to have chiral gapless edge states as well as a gapped bulk, independent of the orientation of the edge. The weak edge states must therefore be stabilized by the appearance of a small mass term. In fact, the same holds for disorder -- the full Hamiltonian belongs to class A, in which the bulk states become localized. Due to the non-zero Chern number, the edge states cannot disappear, and must therefore be stabilized by disorder.

\begin{figure}[tb]
\centering
\subcaptionbox{\label{phase_diag_1a}}
{\includegraphics[width = 0.51 \linewidth]{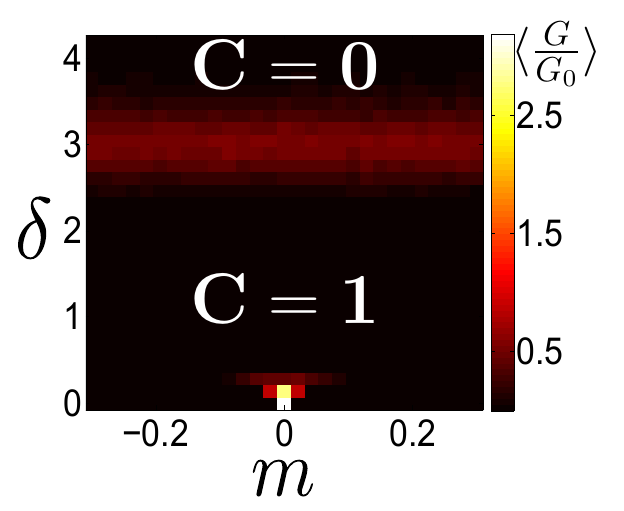}}
\subcaptionbox{\label{4DC_transport}}
{\includegraphics[width = 0.47 \linewidth]{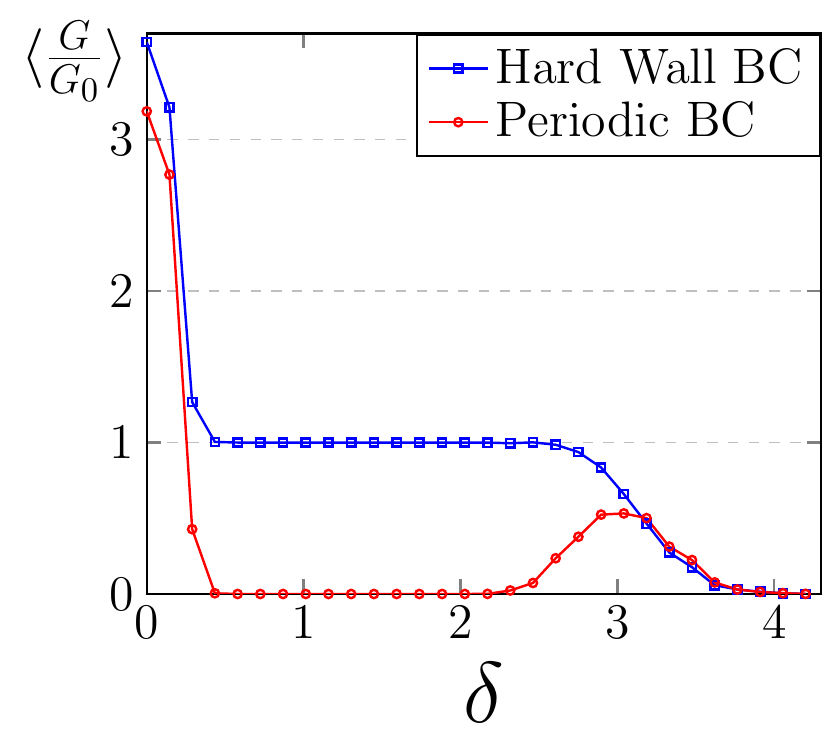}}
\caption{(a) Disorder averaged bulk conductance of \textit{model I} as a function of disorder strength $\delta$ and the gap opening parameter $m$. Chern numbers of the two phases are shown. (b) Average conductance $G/G_0$, for $m=0$, as a function of $\delta$, for periodic (red) and hard wall (blue) boundary conditions, see Ref.~\cite{Parameters1} for simulation parameters. The difference between the two curves is due to edge state conductance.}
\end{figure}

To confirm this expectation, we numerically analyze two-terminal transport in the system. All transport simulations are performed using the Kwant code \cite{kwant}. We discretize the Hamiltonian on a square lattice of $L\times W$ sites, and attach ideal leads in the $x$-direction. This enables us to compute the scattering matrix
\begin{equation}
 S = \begin{pmatrix}
      r & t \\
      t' & r'
     \end{pmatrix},
\end{equation}
which we use to determine the conductance $G/G_0 = {\rm Tr}\,t^\dag t$, $G_0=e^2/h$, in the low bias voltage, low temperature limit. In the $y$-direction, we use either hard-wall boundary conditions (Hall bar geometry), or apply periodic boundary conditions to the states, $\psi(x, 0)=\psi(x, W)$, to access only the bulk contribution to transport (Corbino geometry). Here and in all other models, disorder is introduced as a random variation of the Fermi energy, drawn independently for each lattice site from the uniform distribution $[-\delta, \delta]$.

As seen in \refFig{4DC_transport}, when the disorder strength $\delta$ increases the bulk contribution goes to zero, while the edge contributions become quantized. The phase diagram is obtained by performing transport simulations with periodic boundary conditions and in the presence of a mass term (\refFig{phase_diag_1a}).
Starting from the gapless point, $m=\delta=0$, both the addition of a mass term or disorder drive the system into a Chern insulating phase with $C=1$.

\textit{\label{QHEQHEmodel}Model II --- A gapped Chern insulator coupled to a Chern insulator at its critical point}. In this model, we keep $H_1$ as before, but replace $H_2$ by a Hamiltonian of a quantum Hall state at the transition between two Chern numbers. This Hamiltonian is nothing but the Hamiltonian appearing in Eq.~\ref{AQH} with $\sigma_z\to-\sigma_z$ and with $\mu=2$. Here, the gapped $H_2$ belongs to class A with either Chern number zero or $-1$. Therefore, the full Hamiltonian is a class A Chern insulator with a Chern number changing from $C=0$ to $C=1$. In contrast to model I, here the edge states disappear as disorder is introduced, since the system enters a $C=0$ phase. The phase diagram of this model is shown in appendix A.

\textit{\label{QHBHZmodel}Model III --- A gapped Chern insulator coupled to a quantum spin Hall state at its critical point}. In the previous models, both the gapped and the gapless Hamiltonian were subjected to localization by disorder. Now we choose an $H_2$ that does not get localized by weak disorder. Interestingly, we find that its coupling to the gapped Chern insulator makes it amenable to localization. We set $H_2$ to be the Bernevig-Hughes-Zhang (BHZ) model for the quantum spin Hall effect \cite{BHZModel},
\begin{equation}\label{BHZ}
H_{2}=\begin{pmatrix}
h(\textbf{k}) & \Gamma(\textbf{k}) \\
\Gamma^{\dagger}(\textbf{k}) & h^*(\textbf{-k})
\end{pmatrix}
\end{equation}
with
\begin{align}\label{BHZ+comp}
\begin{split}
h(\textbf{k}) =& \left( M_0+2M_2(1-\cos{k_x}-\cos{k_y} \right)\sigma_z +A\sigma_x\sin{k_x}+ \\
 & \left( C_0+2C_2(1-\cos{k_x}-\cos{k_y} \right)\sigma_0-A\sigma_y\sin{k_y}, \\
\Gamma(\textbf{k}) =& \Delta \left( \sigma_0\sin{k_x} +i\sigma_z\sin{k_y}\right)-i\Delta_0\sigma_y,
\end{split}
\end{align}
where the $\sigma_i$'s act in the subspace of the $E$ and $H$ orbitals of the BHZ model. We choose $M_0$ such that $H_{2}$ is in a metallic region between two non-trivial quantum spin Hall phases (symmetry class AII). For the full model, we take $H_1$ as in \refEq{AQH}, and a simple coupling Hamiltonian,
\begin{equation}
 H_c=t\begin{pmatrix}
      1 & i & 1 & i \\
      1 & i & 1 & i
     \end{pmatrix}.
\end{equation}

\begin{figure}[tb]
\centering
\subcaptionbox{\label{QAH_BHZ_PD}}
{\includegraphics[width = 0.52 \linewidth]{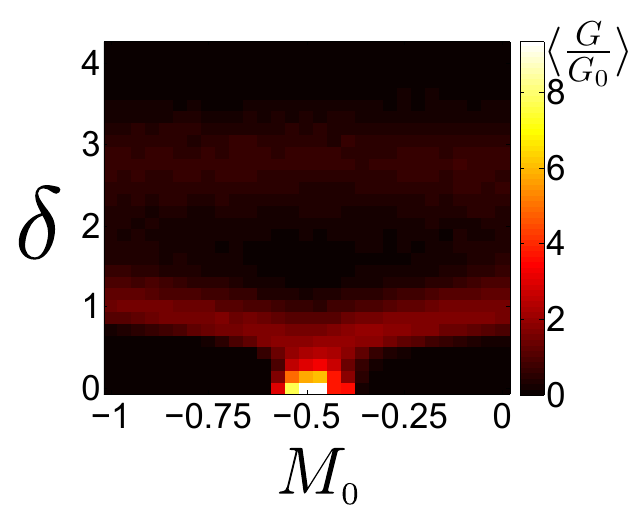}}
\subcaptionbox{\label{bhz_scaling_mid}}
{\includegraphics[width = 0.46 \linewidth]{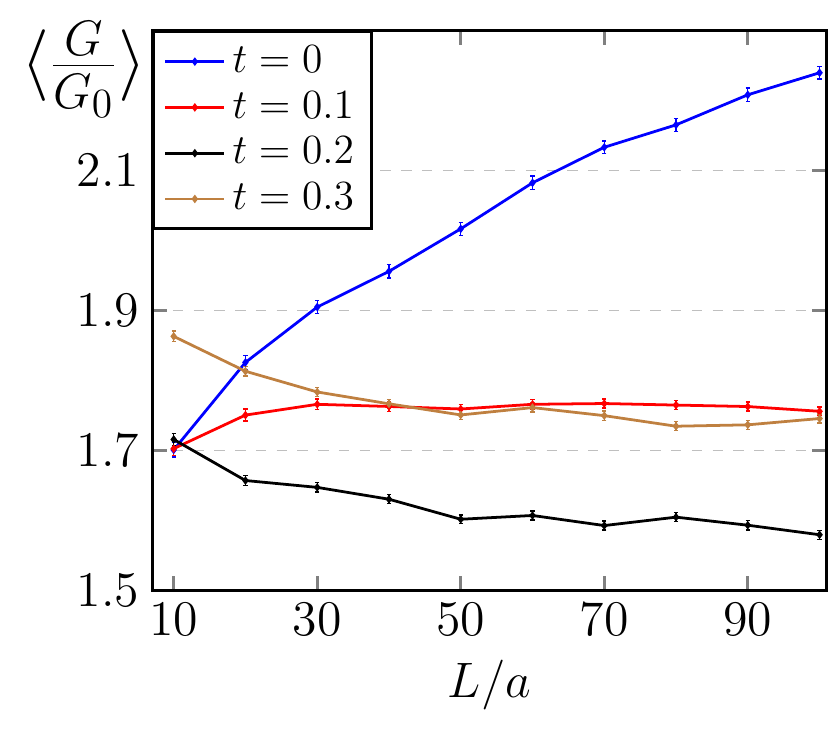}}
\caption{(Color online) (a) Disorder averaged bulk conductance (color scale) of \textit{model III}, as a function of $M_0$ and disorder strength $\delta$. (b) Average bulk conductance for $M_0=-0.75$ and $\delta=0.87$, as a function of the system size $L$ for different coupling strengths.  The conducting region around $\delta=1$ becomes localized as the system size is increased. See \cite{params3} for simulation parameters.}
\end{figure}

Similar to Figs.~\ref{fig:openYX}a and \ref{fig:openYX}b, in the absence of disorder, bulk and edge modes coexist in the spectrum, as shown in appendix B. While $H_{2}$ is time-reversal symmetric, allowing for the existence of metallic phases in the presence of disorder, the coupled model belongs to class A, where weak disorder leads to localization. Seemingly, the phase diagram in the space of $M_0$ and disorder strength $\delta$, depicted in \refFig{QAH_BHZ_PD}, shows a metallic phase at finite disorder strength (close to $\delta=1$), reminiscent of that present in the BHZ model. 
However, a conductance scaling analysis (\refFig{bhz_scaling_mid}) shows that it is only metallic in the decoupled case. When the coupling is turned on, the conductance decreases with system size, showing that the presence of the conducting region is caused by the finite-size of the system. In accordance with \refFig{phase_diag_1a}, as the disorder strength is increased, the bulk states localize leaving behind only the quantized conductance contribution of the edge states. This exemplifies the fact that both the topological and the localization properties depend on the symmetry class of the full Hamiltonian. In the thermodynamic limit, the phase diagram of this model should be identical to that of \textit{model I}.

\textit{\label{DQWmodel}Model IV --- A gapped quantum spin Hall phase coupled to a quantum spin Hall phase at its critical point}.
We now consider both $H_1$ and $H_2$ to be BHZ models, Eqs.~\eqref{BHZ} and \eqref{BHZ+comp}, with different mass terms, $M_{0,1}$ and $M_{0,2}$, respectively. We set $H_1$ to be in a topological phase and $H_2$ in a metallic region.
This model can be directly implemented experimentally, for example with two coupled Hg(Cd)Te quantum wells as proposed in \citeRef{Michetti2012}.
Most directly, this model may be realized in such systems when one of the quantum wells is grown with a critical thickness \cite{BuettnerNature2011} while the other well is chosen to be in a topologically non-trivial phase. Remarkably, the system may be driven to the gapless point (see appendix C) by the application of voltage on front and back gates even when the thickness of the two wells does not conform to this requirement. 

\begin{figure}[tb]
\centering
\subcaptionbox{\label{FigBHZBHZpd}}
{\includegraphics[width = 0.51 \linewidth]{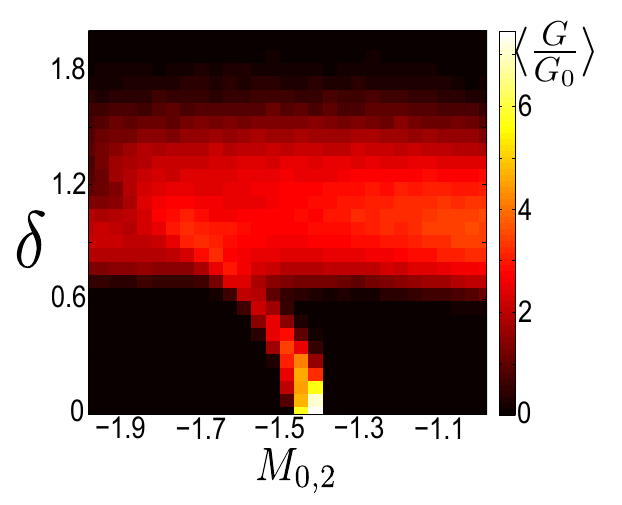}}
\subcaptionbox{\label{bhz_bhz_scaling}}
{\includegraphics[width = 0.46 \linewidth]{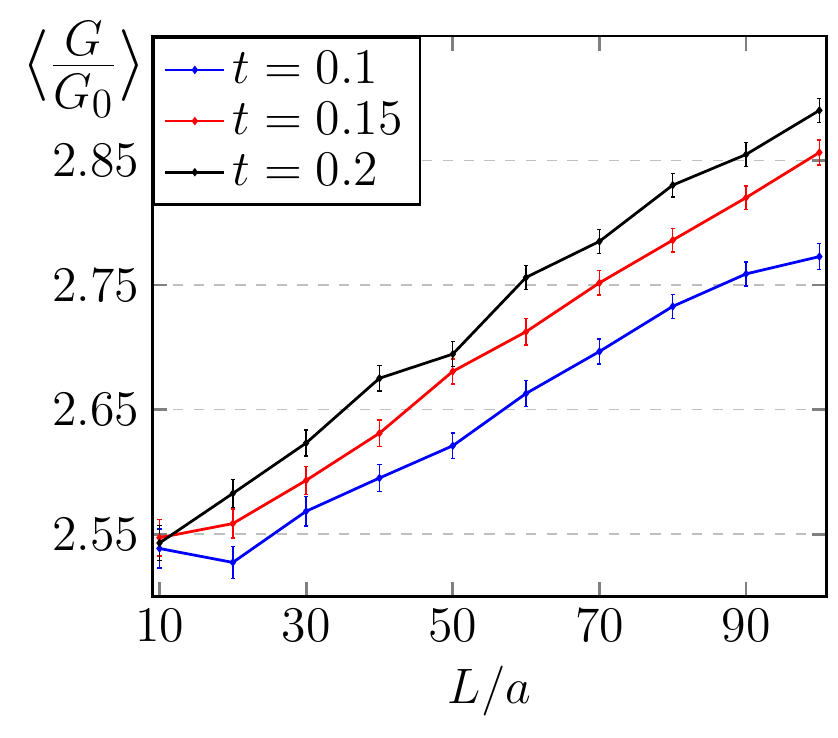}}
\caption{(Color online) (a) Disorder averaged bulk conductance of two coupled BHZ models, as a function of disorder strength $\delta$ and $M_{0,2}$. (b) Average bulk conductance for $\delta=1$ and $M_{0,2}=-1.47$, as a function of the system size $L$ for different couplings $t$. See \cite{BHZBHZparam222} for simulation parameters.}
\end{figure}

We choose a coupling Hamiltonian $H_c = t \left(\sigma_0 + \sigma_z\right)\tau_0$, found in Ref.~\cite{Michetti2012} to describe the experimentally accessible parameter regime. The Pauli matrices $\sigma$ describe the space of the $E$ and $H$ orbitals, and $\tau$ parametrize the spin degree of freedom.
In the absence of disorder (appendix C), we find features similar to the previous models, the system simultaneously hosting gapless modes in the bulk and on the edge. Like in \textit{model III}, when  $M_{0,2}$ is changed the bulk becomes insulating, but this time it is characterized by a different invariant, belonging to $\mathbb{Z}_2$ instead of the $\mathbb{Z}$ valued Chern number. The disordered case also shows a behavior different from the previous models. Since the full Hamiltonian belongs to class AII, weak anti-localization leads to the formation of a metallic phase at finite disorder strength, see Fig.~\ref{FigBHZBHZpd}. This is confirmed by the scaling analysis of Fig.~\ref{bhz_bhz_scaling}, showing that the conductance increases with system size.

\textit{\label{Summary}Summary ---}
We have suggested a route to theoretically and experimentally realize unusual two dimensional topological phases that simultaneously host gapless modes in the bulk and on the edges. We find that the modes on the edge exhibit a peculiar structure that depends on the boundary orientation. Additionally, we find that the behavior of these phases in the presence of disorder can be extracted from the clean limit by analyzing their topological properties in the presence of an infinitesimal bulk gap.

\textit{Acknowledgments ---}
Financial support by the DFG (German-Japanese research unit "Topotronics"; priority program SPP 1666 "Topological insulators"), the Helmholtz Foundation (VITI), and the ENB Graduate School on "Topological Insulators" is gratefully acknowledged by TP and BT. TP wants to thank P.~Michetti for interesting discussions and the Weizmann Institute for hospitality. YB, ICF, and AS gratefully acknowledge support from the European Research Council under the European Union's Seventh Framework Programme (FP7/2007-2013) / ERC Project MUNATOP, the US-Israel Binational Science Foundation and the Minerva Foundation.

\section{Appendix}
\subsection{A. Model \textit{II}}
The second model presented in the main text consists of two coupled quantum anomalous Hall systems (QAH), with the Hamiltonian

\begin{equation}\label{QAH_FULL} 
H_{\text{II}}=\begin{pmatrix}
H_1 & t(\sigma_0+\sigma_x) \\
t(\sigma_0+\sigma_x) & H_2
\end{pmatrix},
\end{equation}
where
\begin{align}
H_{1,2}=&\epsilon(\textbf{k})\pm (\mu_{1,2}-\cos{k_x}-\cos{k_y})\sigma_z\\ \nonumber
&+v_1(\sigma_x\sin{k_x}+\sigma_y\sin{k_y}),
\end{align}
and $\epsilon(\textbf{k})=t_1(1-\cos{k_x}-\cos{k_y})\sigma_0$.

We choose $\mu_1=1$ such that $H_1$ is in a non-trivial phase with $C=1$. For $\mu_2=2$, $H_2$ is at a transition between two Chern insulating phases with $C=-1$ for $\mu_2<2$, and $C=0$ for $\mu_2>2$. As such, the combined system is trivial for $\mu_2<2$ and has $C=-1$ for $\mu_2>2$.

Similar to the other models, the Hamiltonian \eqref{QAH_FULL} was discretized on a square lattice of $L\times W$ sites (lattice constant $a=1$), and disorder was introduced as a random change of the on site Fermi energy, chosen independently for each lattice site from the symmetric uniform distribution $[-\delta, \delta]$. We have used $L\times W=80\times80$, $t_1=t=0.2$, and $v_1=1$.

\begin{figure}
\begin{center}
\subcaptionbox{ \label{QAH_QAH_PD}}%
{\includegraphics[width=0.51 \linewidth]{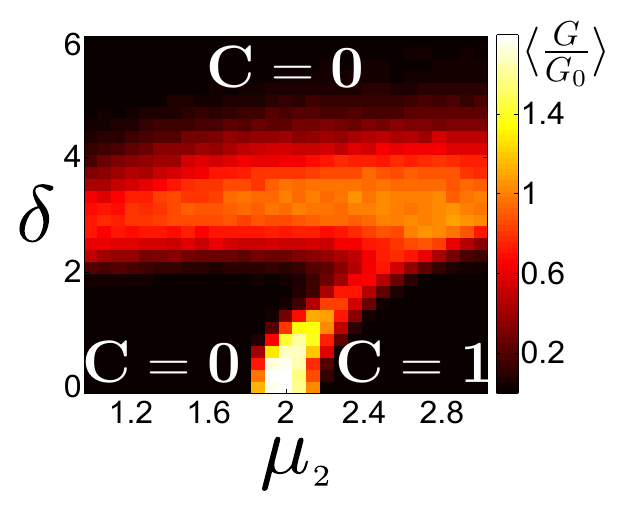}}
\subcaptionbox{\label{QAH_QAH_sc}}
{\includegraphics[width = 0.45 \linewidth]{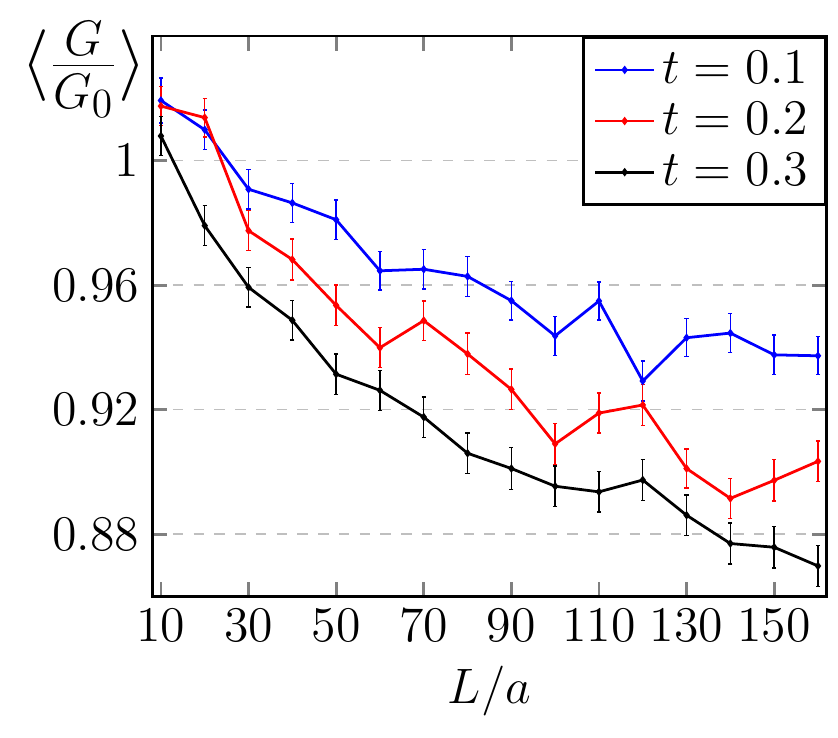}}
\caption{ (Color online) (a) Average bulk conductance of Eq.~(\ref{QAH_FULL}) as a function of disorder strength $\delta$ and $\mu_2$. Each point was obtained by averaging over 100 disorder realizations. (b) Average bulk conductance for $\mu_2=2$ and $\delta=3.1$, as a function of the system size $L$ for different coupling strengths $t$.}
\end{center} 
\end{figure}

The phase diagram as a function of disorder strength $\delta$ and $\mu_2$ is shown in \refFig{QAH_QAH_PD}. The bulk insulating phases are separated by a critical line across which the edge states localize, when the topological invariant changes from $C=1$ to $C=0$. In addition, there is a conducting region throughout the plotted $\mu_2$ range close to $\delta=3$, similar to that in Fig.~3. This feature is a finite-size effect which appears due to the different gaps of $H_1$ and $H_2$. As in \textit{model III}, the gap of $H_1$ is kept constant while that of $H_2$ is closed and reopened when it undergoes a topological phase transition. Therefore, increasing the strength of disorder causes $H_1$ and $H_2$ to Anderson localize at different values of $\delta$, leading to a finite-size effect which is more pronounced for a larger difference between their gaps. The scaling analysis of \refFig{QAH_QAH_sc} confirms the nature of this conducting region, showing that bulk conductance decreases with increasing system size. In the infinite system size limit, starting from the critical point of the clean system, $\mu_2=\delta=0$, both the bulk and edge conductance would go to zero for $\mu_2=0$ and any $\delta > 0$.

\subsection{B. Model \textit{III}}
\begin{figure}
\includegraphics[width=0.48 \linewidth]{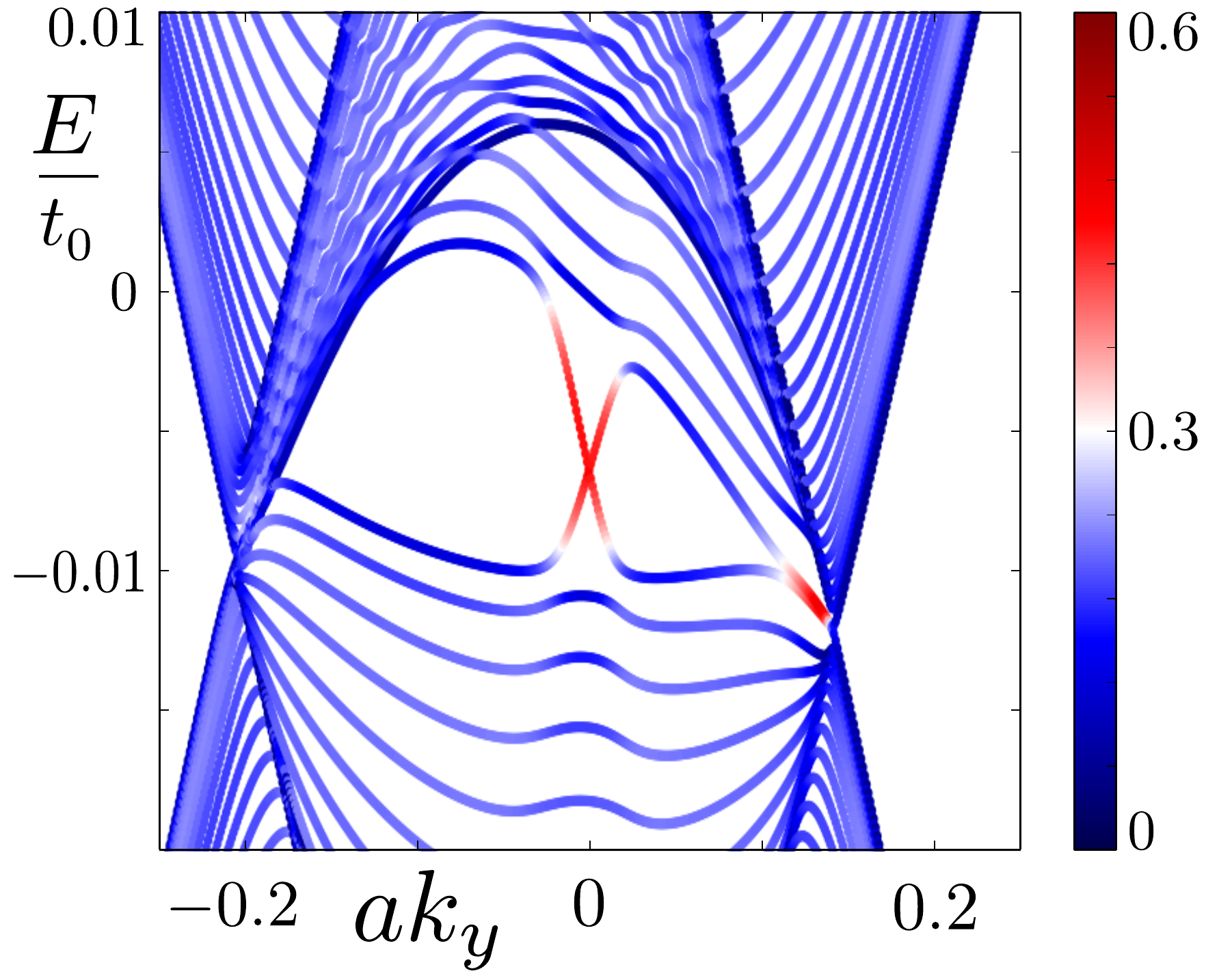}
\includegraphics[width = 0.48 \linewidth]{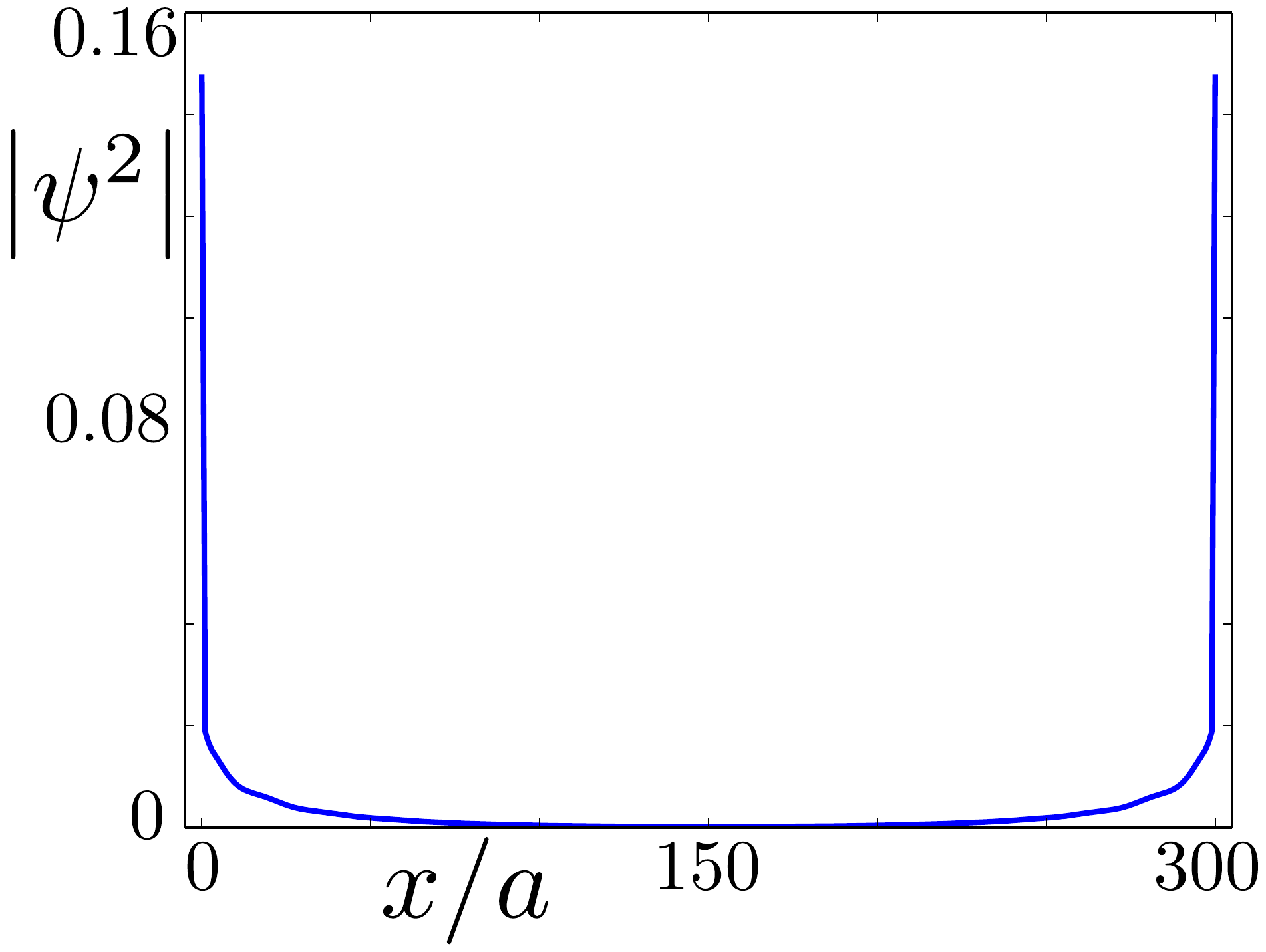}
\caption{ (Color online) Left panel: Band structure of \textit{model III} with open boundary conditions in the $x$-direction. The color scale shows the weight of the state on the first and last $10\%$ of lattice sites. Right panel: Intensity as a function of position for the edge states at $k_y=0$ in the left panel. \label{fig:bhzqh_bands_wf}}
\end{figure}

Coupling a system in a quantum Hall phase to a BHZ model in a metallic phase leads to a spectrum similar to that shown in the main text as depicted in Fig.~1. Bulk and edge states coexist at the Fermi level and overlap in momentum, causing them to hybridize (Fig.~\ref{fig:bhzqh_bands_wf}). Even though they no longer decay exponentially into the bulk, their wave functions are still visibly peaked at the edges of the system. We use $M_0=-0.49$, $t=0.1$, keeping all other parameters unchanged compared to the main text.

\subsection{C. Model \textit{IV}}
\begin{figure}[t]
\includegraphics[width=0.48 \linewidth]{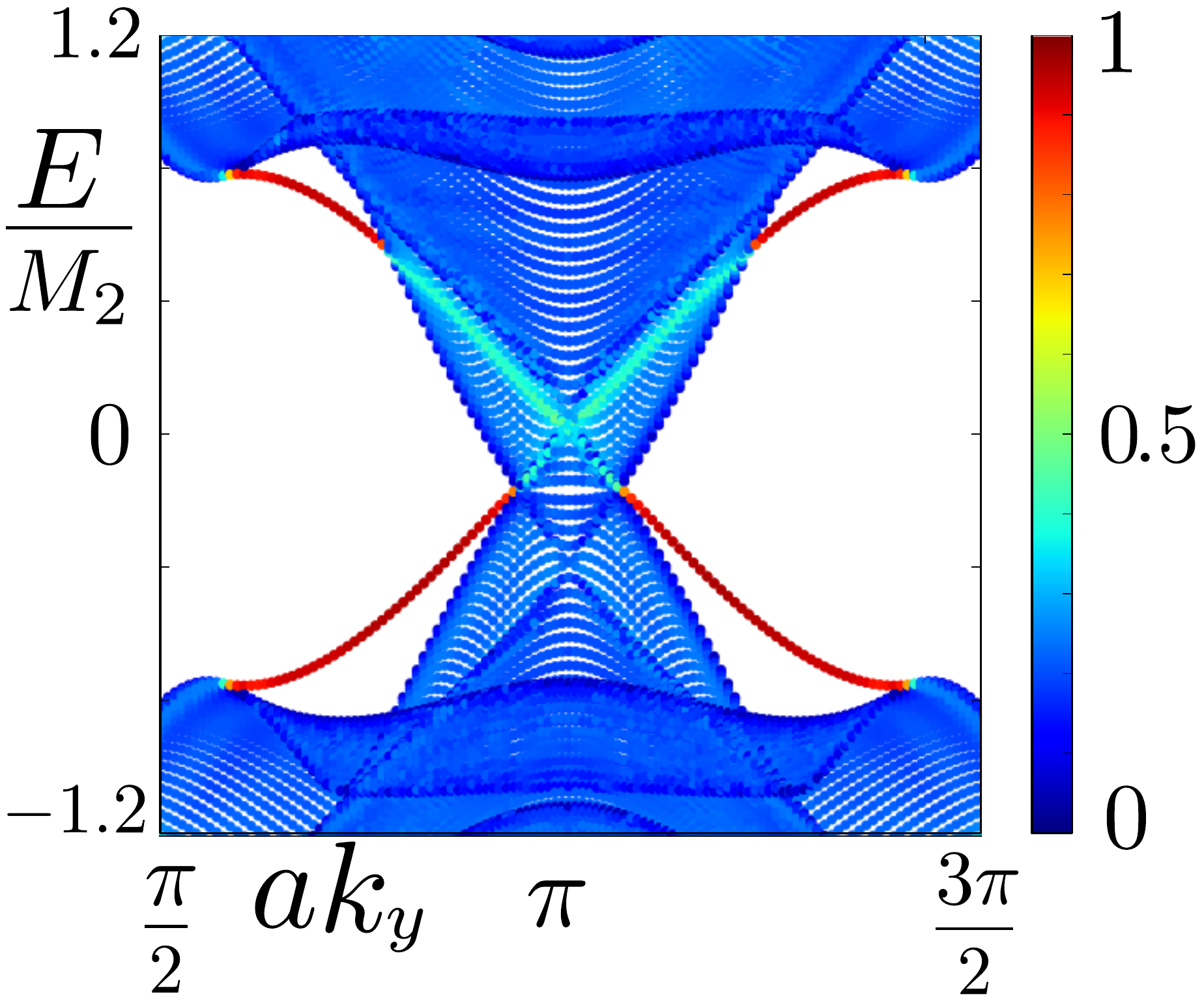}
\includegraphics[width = 0.48 \linewidth]{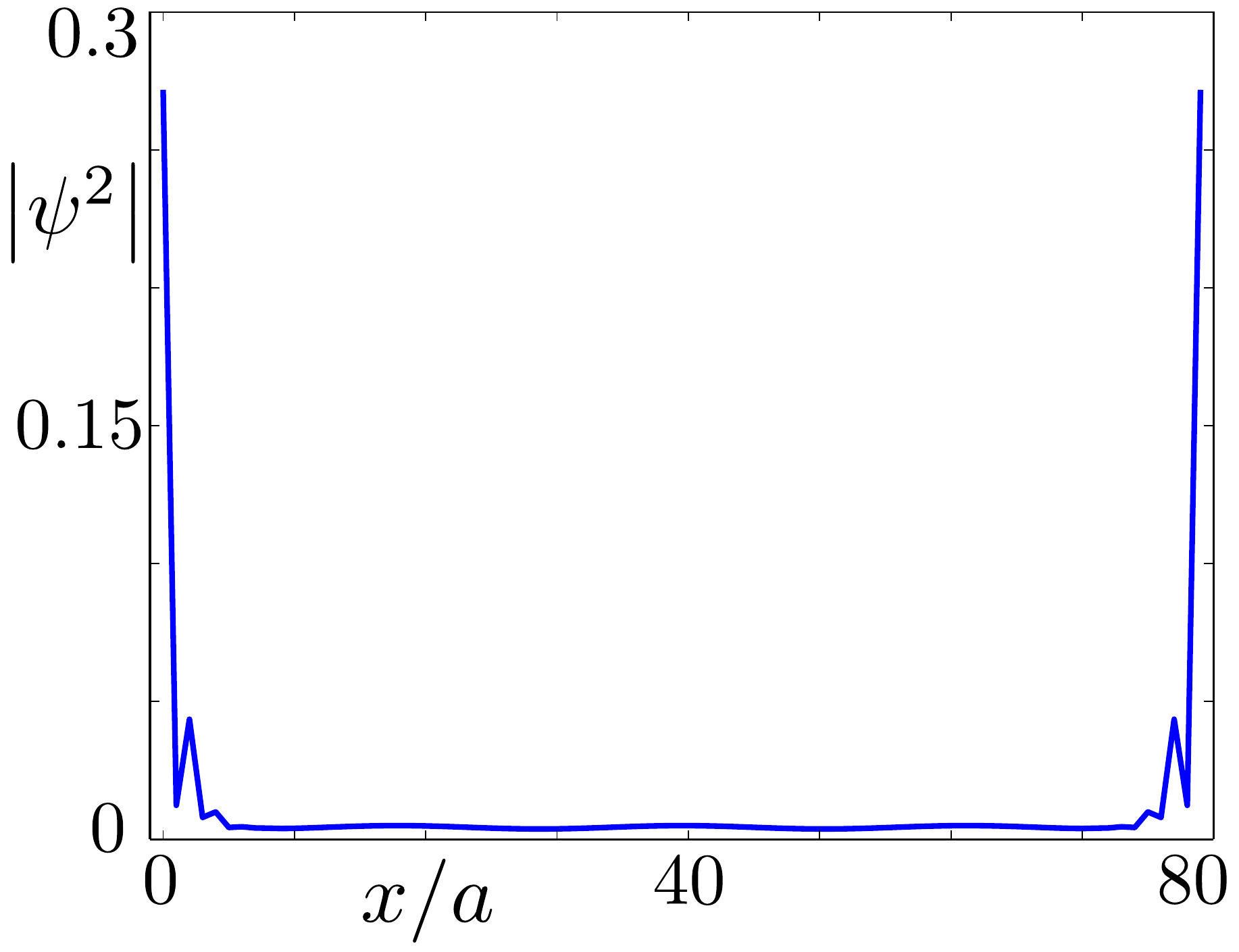}
\caption{ (Color online) Left panel: Band structure of \textit{model IV} with open boundary conditions in the $x$-direction, showing hybridized bulk and edge states. The color scale shows the relative intensity of each state on the first and last $10\%$ of lattice sites. Right panel: Intensity as a function of position for the edge states located at $k_y=\pi$ in the left panel.\label{fig:bhzbhz_bands_wf}}
\end{figure}

As in the other models, Fermi level bulk and edge states can coexist in two coupled BHZ Hamiltonians. For some energies, edge and bulk states overlap, and therefore hybridize. Nevertheless, the spatial profile of the hybridized wave functions shows clear peaks at the edges of the system. We illustrate this observation in Fig.~\ref{fig:bhzbhz_bands_wf}, using $M_{0,1}=-1$, $M_{0,2}=-1.45$, and keeping all other parameters as in the main text.

\newlength{\smallLength}

\begin{figure}
\centering
\includegraphics[width = 0.6 \linewidth]{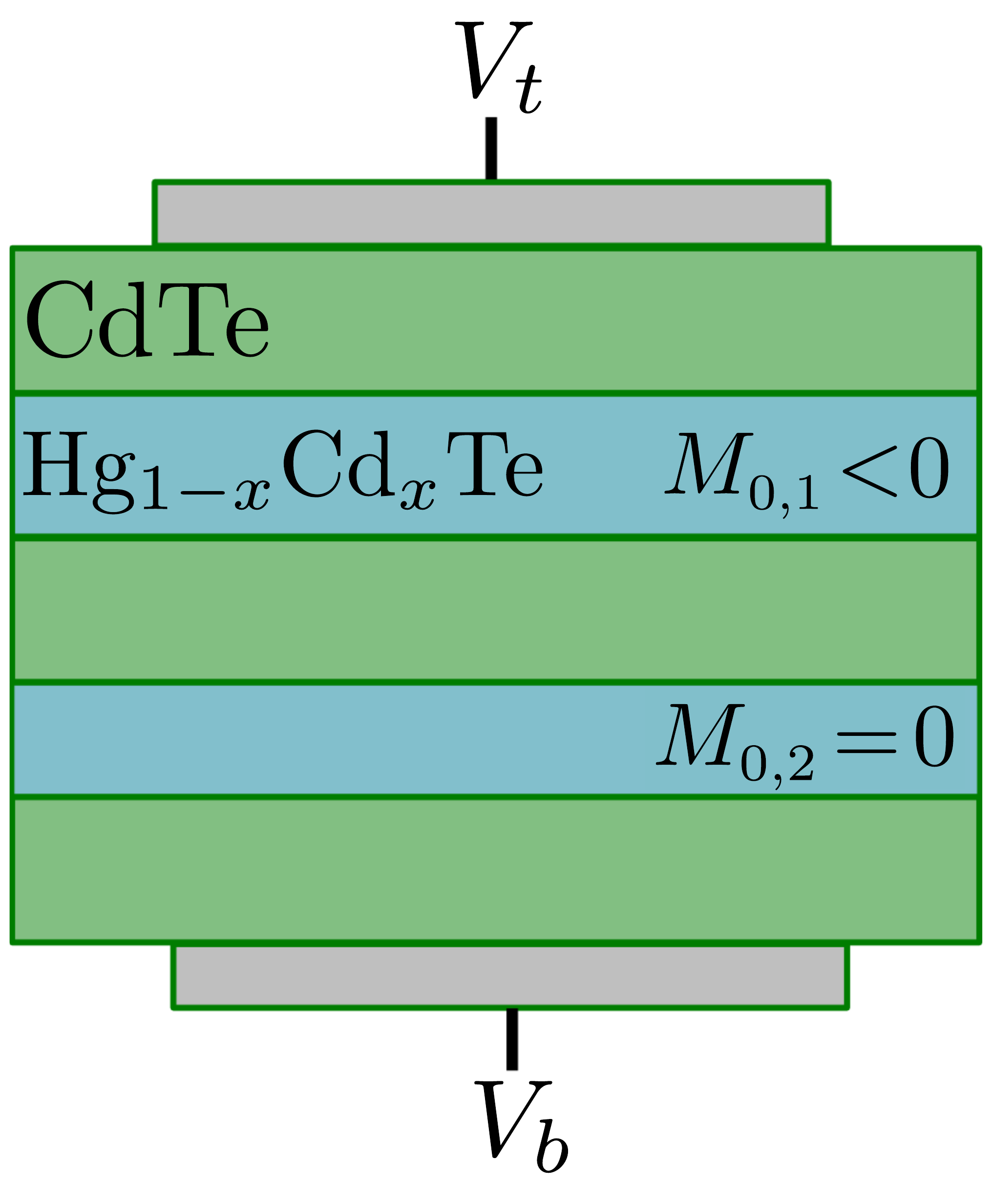}
\caption{\label{figSetupDoubleQWellAppendix}} Schematic setup of a double quantum well structure in Hg(Cd)Te, realizing a gapless topological phase.
\end{figure}

\textit{Model IV} is realizable in a double HgTe quantum well structure in a CdTe matrix (\refFig{figSetupDoubleQWellAppendix}), a setup proposed in \citeRef{Michetti2012}. One of the quantum wells should be grown with the critical thickness of $6.3$nm, which has been experimentally realized in \citeRef{koning}, while the other is chosen to be topologically non-trivial, with a width of, e.g., $7$nm.
Both are described by the BHZ model of \refEq{BHZ}, where the parameters are adjusted to the thickness of the respective quantum well. For this, we use the more experimentally relevant estimates of the parameters of Refs.\onlinecite{RotheReinthalerLiuMolenkampZhangHankiewicz2010FingerprintOfDifferentSpinOrbitTermsForSpinTransportInHgTeQuantumWells} and \onlinecite{BuettnerNature2011}, summarized in Table 
\ref{tabParametersUsedAppendix}.
\begin{table}[tb]
\begin{tabular}{cccccc}
	thickness	&	$A$/eV	&	$M_2$/eV	&	$C_0$/eV&	$C_2$/eV	&	$M_0$/eV
\\
\hline
$7.0$ nm &	$0.365/a_c$&	$-0.706/a_c^2$ & 0	&	$-0.532/
  a_c^2$& $0.01009$ 
\\
$6.3$ nm	&	$-0.373/a_c$ & $-0.857/a_c^2$	& $0$	& $-0.682/a_c^2$	& $0$
\end{tabular}
\caption{\label{tabParametersUsedAppendix}Parameters used for the non-trivial quantum well (top row), taken from \citeRef{RotheReinthalerLiuMolenkampZhangHankiewicz2010FingerprintOfDifferentSpinOrbitTermsForSpinTransportInHgTeQuantumWells}, and the gapless one (bottom row), from \citeRef{BuettnerNature2011}. We choose an effective lattice constant $a_c = 0.64$nm, and absorb the chemical potentials, parametrized by $C_0$, into the gate voltages, setting them to zero (see text).}
\end{table}

For clarity, we neglect off-diagonal terms in the BHZ models, such as bulk and structural inversion asymmetry,
\cite{LiuHugesQiQangZhang2008QuantumSpinHallEffectInInvertedTypeIISC, RotheReinthalerLiuMolenkampZhangHankiewicz2010FingerprintOfDifferentSpinOrbitTermsForSpinTransportInHgTeQuantumWells, BHZModel} since they vanish at the $\Gamma$ point, where the bulk gap closes.
The two quantum wells are coupled by the term introduced in \citeRef{Michetti2012}, which fulfills the following properties:
\begin{enumerate}
\item the overlap between the $E1$ states of the quantum wells is much stronger than the overlap between the $H1$ states~\footnote{The underlying reason is the interfacial character of the $E1$ quantum well bands opposed to the intra well character of the $H1$ bands.}, and
\item at the $\Gamma$ point, the overlap between the $E1$ and $H1$ states of different quantum wells vanishes due to rotational symmetry along the axis of crystal growth~\footnote{
The introduction of a boundary and the lattice structure of the material system in fact break this symmetry. However, the former symmetry violation is a finite size effect and the latter becomes irrelevant close to the $\Gamma$ point}.
\end{enumerate} 
These properties guarantee a simple structure of the inter layer coupling at the $\Gamma$ point,
\begin{align}\label{eq:coupling_H}
H_\text{coup}(0) = t \left( \sigma_0 + \sigma_z \right) \tau_0 \rho_x,
\end{align}
where $\sigma$, $\tau$ and $\rho$ are Pauli matrices describing the band-index, time-reversal and layer degree of freedom, respectively.

The value of $M_{0,2}$ at which the topological phase transition occurs changes as a function of the coupling strength $t$, such that the well of critical thickness becomes a gapped, topologically trivial system when $t > 0$. However, for any coupling Hamiltonian that preserves time-reversal symmetry there exists a value of $M_{0,2}$ at which the combined system is a gapless topological insulator.

Rather than tuning the width of the quantum wells to reach this point, the bulk gap my be closed by shifting their chemical potentials relative to each other. This can be done by applying different gate voltages on top and bottom gates, $V_t$ and $V_b$,  as shown in \refFig{figSetupDoubleQWellAppendix}. The combined system then becomes gapless when the top of the valence band in one well and the bottom of the conduction band in the second well touch each other.

\begin{figure}[t]
\subcaptionbox{ \label{VCN}}%
{\includegraphics[width=0.3 \linewidth]{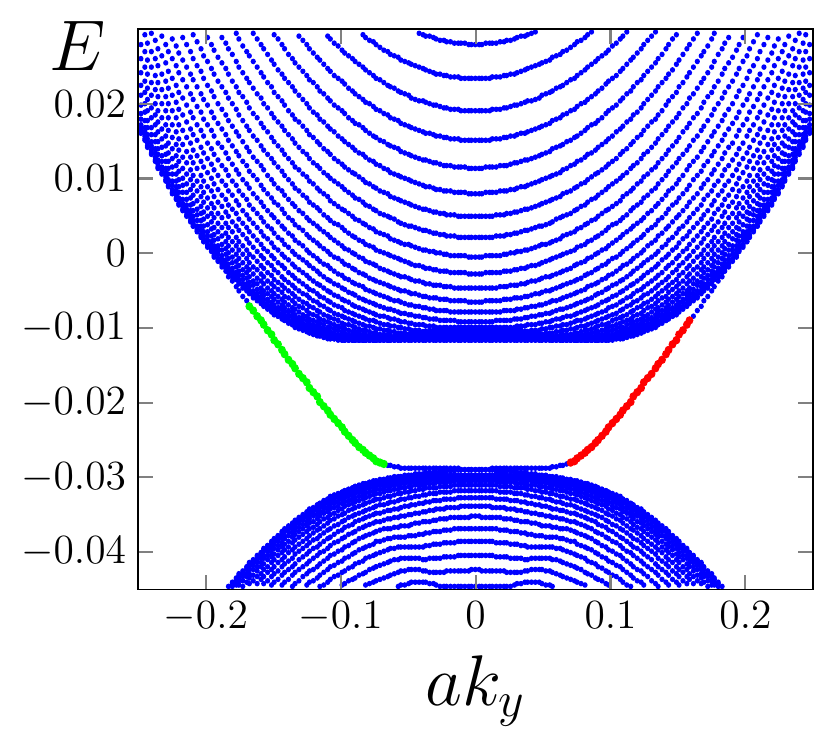}}
\subcaptionbox{ \label{VC}}%
{\includegraphics[width = 0.3 \linewidth]{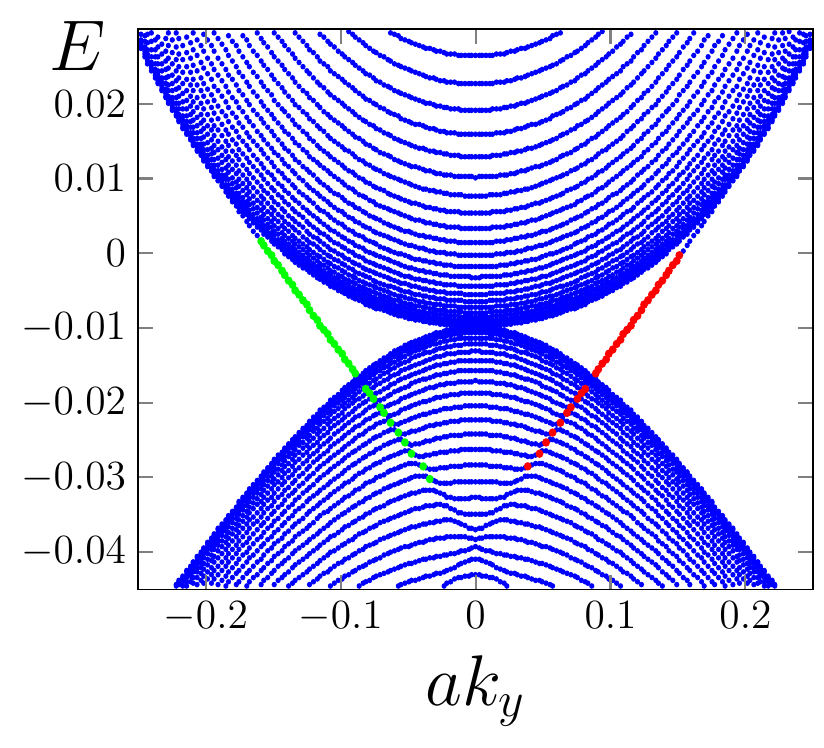}}
\subcaptionbox{ \label{VCP}}%
{\includegraphics[width=0.3 \linewidth]{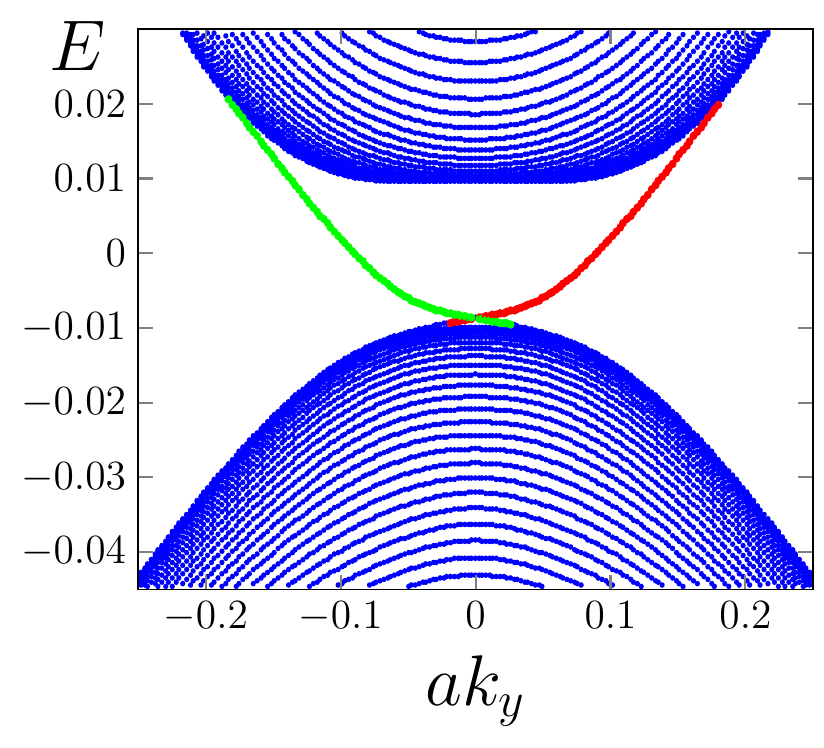}}
\caption{ (Color online)  Band structure of \textit{model IV} with open boundary conditions in the $x$-direction. $M_{0,1}=-M_{0,2}=0.01$ and the coupling between the layers is $t=0.1$. Other parameters are as in \refTab{tabParametersUsedAppendix}. The band structure is calculated for three different gate voltage configurations such that (a) $\Delta V= V_c-0.02$ (b) $\Delta V= V_c$ (c) $\Delta V= V_c+0.02$. The red and green points denote left/right moving edge states, while the blue points denote bulk states. \label{Gatevol_effect}}
\end{figure}

Due to the coupling between the layers a gap is reopened when the two bands overlap, with both gapped phases being topologically non-trivial $Z_2$ insulators. 
For the coupling in \refEq{eq:coupling_H}, there is a critical voltage difference, $\Delta V\equiv V_c$, at which the bulk gap closes. Generally, the critical voltage, $V_c$, depends on the bulk gaps and chemical potentials of the uncoupled wells, and also on the coupling between the layers. However, for the experimentally accessible case where $|M_{0,1}-M_{0,2}|\ll t$, the critical voltage is $V_c = (M_{0,1}-M_{0,2})-(C_{0,1}-C_{0,2})$, independently of $t$.  
Experimentally, the bulk gap closing should appear as a region of increased conductance through the sample, as a function of $V_t$ and $V_b$.
We show the band structure of such a system for three different gate voltages in \refFig{Gatevol_effect}. Here, $M_{0,1}$ and $M_{0,2}$ are chosen such the top layer is in a topological insulating phase while the bottom layer is in a trivial insulating phase.

%


\begin{thebibliography}{25}
\bibitem [{Par()}]{Parameters1}%
  \BibitemOpen
  \href@noop {} {}\bibinfo {note} {The lattice size is $L\times W=80\times 80$,
  and Hamiltonian parameters are $\mu=v_1=1$, $t_1=0$, $v_2=0.2$, and $t=0.1$.
  Each point is obtained by averaging over $100$ independent realizations of
  disorder.}\BibitemShut {Stop}%
\end{thebibliography}

\end{document}